\documentclass[aps,pra,groupedaddress,twocolumn,longbibliography]{revtex4-1}

\usepackage{graphicx, subfigure}
\usepackage{amsmath}
\usepackage{amsfonts}
\usepackage{amssymb}
\usepackage{dsfont}
\usepackage{bbm}

\usepackage{subfigure}

\usepackage{color}

\newcommand{\beqs}{\begin{equation*}}
\newcommand{\eeqs}{\end{equation*}} 

\newcommand{\beq}{\begin{equation}}
\newcommand{\eeq}{\end{equation}} 

\newcommand{\bea}{\begin{eqnarray}}
\newcommand{\eea}{\end{eqnarray}}

\begin{document}

\title{Time-resolved pump-probe signals of a continuously driven quantum dot affected by phonons}

\author{D.~E.~Reiter}
\email[]{doris.reiter@uni-muenster.de}

\affiliation{Institut f\"ur Festk\"orpertheorie, Universit\"at M\"unster, Wilhelm-Klemm-Str. 10, 48149 M\"{u}nster, Germany}

\date{\today}

\begin{abstract}
The interaction of a light field with a quantum-mechanical system can be studied in an optically controlled semiconductor quantum dot. When driven by a continuous light field switched on instantaneously, the quantum dot occupation performs Rabi oscillations. Unlike an atomic system, the quantum dot is coupled to phonons, which leads to a damping of the Rabi oscillations. Here we model the time-resolved probe spectra to monitor these dynamics and study the influence of phonons on the spectra. The spectra consist of up to three peaks, similar to the Mollow-triplet known from quantum optics. We develop analytical equations within a rate equation model and show that they agree excellently with a numerical solution using a well established correlation expansion approach. 
\end{abstract}

\maketitle

\section{Introduction} \label{sec:intro}
A semiconductor quantum dot (QD) coupled to a light field is at the heart of using semiconductor nanostructures for ultra fast applications in quantum information technology. The optical control of a QD allows to prepare selected states \cite{ramsay2010are,reiter2014the} and in return a QD can be used as single or entangled photon source \cite{press2007pho,dousse2010ult}. In contrast to an atom, a QD is always embedded in a solid state matrix and coupled to its lattice vibrations, i.e., the phonons. For continuous excitation the interaction with phonons leads to a damping of the Rabi oscillations accompanied by phonon wave packet emission \cite{glassl2011lon,wigger2014ene}, while for pulsed excitation phonons can either deteriorate or assisted the preparation \cite{reiter2014the}. Nowadays it is possible to embed a QD in an optical microcavity, e.g. a micropillar \cite{reitzenstein2010qua,fischer2016sel}, where the strong coupling regime can be reached. By using a QD in a microcavity it became possible to experimentally observe the Mollow-triplet \cite{ulhaq2010lin,ulhaq2013det,unsleber2015obs,fischer2016sel}; an effect well known from quantum optics \cite{gerry2005int}. Also for these systems phonons play a crucial role \cite{ulhaq2010lin,roy2011pho,wei2014tem,roy2015qua,das2013col,hopfmann2015com,nazir2016mod}.

In this paper, we will study the time-resolved optical signals resulting from a QD which is driven by a continuous light field switched on instantaneously. In this case the QD occupation performs damped Rabi oscillations due to the interaction with phonons. To be specific, we simulate a pump-probe set-up where the switched-on continuous light is regarded as the pump field and the system is probed by an ultrafast pulse. To discriminate the signal of the pump pulse from the probe pulse either a spatial separation via the propagation direction can be performed \cite{mukamel2000mul}, while in the case of a single QD a frequency modulation in combination with a heterodyne detection technique can be used as established in four wave mixing (FWM) experiments \cite{kasprzak2013coh,mermillod2016dyn}. The probe spectrum reflects the eigenenergies of the coupled QD-light system forming the Mollow triplet. Note that the Mollow triplet already arises from the interaction with a classical field \cite{del2010mol,mollow1969pow}, although it is typically described in a quantum picture of the fluorescence signal \cite{gerry2005int}. In the dynamical picture, the occupation of the QD changes with time, which is reflected in the strength of the three peaks, whose amplitudes correspondingly oscillate in time. When the Rabi oscillations become damped due to the interaction with phonons, the center peak vanishes, yielding a spectrum consisting of only two peak with opposite signs. Hence, in time-resolved optical signals both the Rabi oscillations and the phonon dynamics should be clearly visible. 

We will develop analytical equations for the dynamics and spectra using a rate equation approach in the eigenbasis of the coupled QD-light system, i.e., the dressed state basis. While in the bare state basis the phonon interaction is of pure dephasing type, in the dressed state basis the effect of phonons can be described by a decay rate yielding transitions between the states. We will confirm the validity of the analytical model by comparing the dynamics and the spectra with the results from a numerical solution using a correlation expansion approach, which is a well established method to solve the coupled QD-light-phonon dynamics \cite{krugel2005the,glassl2011lon,reiter2014the}. 

\section{Theoretical background}
\label{sec:theo}
\subsection{Hamiltonian in the bare and dressed state basis}
\label{theo:hamil}

\begin{figure}[t]
\includegraphics[width=\columnwidth]{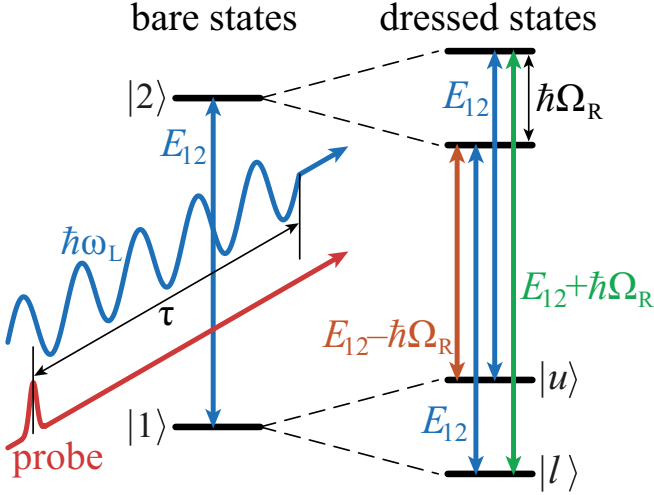}
\caption{(Color online) (left) Sketch of the bare states including the light field and (right) the dressed states. Transitions are indicated by the arrows. \label{fig:scheme}}
\end{figure}

For a strongly confined QD the electronic states can be described as a two-level system as long as the QD is excited by circularly polarized light and the fine structure splitting can be neglected. This two-level system is driven by an external classical light field and we take into account the phonon-interaction. The corresponding Hamiltonian is separated into two parts with
\beq 
	H = H_ {QD-L} + H_{pho},
\eeq
where  $H_{QD-L}$ accounts for the QD system including the light field coupling and $H_{pho}$ contains the phonon part. The QD two-level system consists of the ground state $|1\rangle$ and the exciton state $|2\rangle$, which have the energy difference $E_{12}$. The light field coupling is described semiclassically in the usual dipole and rotating wave approximation. Assuming that the energy of the light field $\hbar\omega_L$ is resonant with the QD transition, $\hbar\omega_L=E_{12}$, our Hamiltonian in the rotating frame is given by
\beqs
	H_{QD-L} = - \hbar \Omega(t) ( |1\rangle \langle 2| +   |2\rangle \langle 1|),
\eeqs
where $\Omega(t)$ is the envelope of the light field. In our case, we consider an excitation with a continuous light field with the Rabi frequency $\Omega_R$ that is switched on instantaneously at $t=-\tau$. To calculate the optical spectra, we add an ultrafast probe pulse with the strength $\Omega_p$ at time $t=0$, such that the envelope of the light field is
\beqs 
	\Omega(t) = \frac{\Omega_R}{2} \Theta(t+\tau) + \Omega_p(t).
\eeqs
In the numerical solution the probe pulse is assumed to be Gaussian, while in the analytical solution the probe pulse is approximated by a $\delta$-function. Neglecting the probe pulse, the Hamiltonian $H_{QD-L}$ can be easily diagonalized leading to the dressed state Hamiltonian 
\beq
	H^{DS}_{QD-L} = - \hbar \frac{\Omega_R}{2}  |l\rangle \langle l| + \hbar \frac{\Omega_R}{2}  |u\rangle \langle u|
\eeq
with the lower dressed state $|l\rangle$ and the upper dressed state $|u \rangle$. The lower and upper dressed states are split by the Rabi energy $\hbar\Omega_R$. The transformation of the density matrix between the bare states and the dressed states can be done by the matrix given by the eigenvectors $|l\rangle=\frac{1}{\sqrt{2}}(|1\rangle+|2\rangle)$ and $|u\rangle=\frac{1}{\sqrt{2}}(|1\rangle-|2\rangle)$ with
\beq S = 
		\begin{pmatrix} 
			\frac{1}{\sqrt{2}} & \frac{1}{\sqrt{2}} \\ 
			-\frac{1}{\sqrt{2}} & \frac{1}{\sqrt{2}}
		\end{pmatrix} 
		\quad \mbox{and} \quad
	 S^{\dagger} = 
		\begin{pmatrix} 
			\frac{1}{\sqrt{2}} & -\frac{1}{\sqrt{2}} \\ 
			\frac{1}{\sqrt{2}} & \frac{1}{\sqrt{2}}
		\end{pmatrix}.	
\eeq
A sketch of the bare and dressed states is given in Fig.~\ref{fig:scheme}. The optical transitions in this system are indicated by arrows. Note that in systems with a periodic driving the energies are quasi-energies modulo $\hbar\omega_L$ \cite{sambe1973ste,harbich1981ele}. Therefore in the dressed state picture four levels are shown in Fig.~\ref{fig:scheme}. A similar picture is obtained using a quantum description of the light field, where the dressed states are combinations of QD and photon states \cite{gerry2005int}.

For the phonons we consider the pure dephasing coupling to acoustic phonons \cite{krummheuer2002the}, which is known to have the largest impact on QD dynamics \cite{ramsay2010dam,reiter2014the}. The phonon part of the Hamiltonian reads
\beq
	H_{pho} = \sum\limits_{\mathbf q} \hbar \omega_{\mathbf q} b^{\dagger}_{\mathbf q} b^{}_{\mathbf q} +  \sum\limits_{\mathbf q} \hbar g_{\mathbf q} (b^{\dagger}_{\mathbf q} +  b^{}_{\mathbf q} ) |2\rangle \langle 2| 
\eeq
with the phonon creation/annihilation operators $b^{\dagger}_{\mathbf q}/b^{}_{\mathbf q}$ with wave vector ${\mathbf q}$. The dispersion is $\omega_{\mathbf q} = c_sq$, $c_s$ being the speed of sound, and the coupling matrix element is denoted by $ g_{\mathbf q}$. The coupling matrix element is calculated using a harmonic oscillator confinement potential for the QD \cite{krummheuer2002the} and its strength depends on the QD size and material parameters. It depends non-monotonically on the energy and the coupling strength is quantified by the spectral density \cite{wigger2014ene,reiter2014the} 
\[
 J(\omega) = \sum_{\mathbf q} |g_{\mathbf q}|^2 \delta(\omega-\omega_{\mathbf q}).
\]
One example of a spectral density is shown in Fig.~\ref{fig:time}~b), where we have used GaAs material parameters and a dot size of 3 nm. The spectral density shows a clear non-monotonic behavior with a maximum at a few meV. 

When transformed into the dressed state basis, the phonon coupling transforms to
\beq
H^{DS}_{pho} = \sum\limits_{\mathbf q} \hbar \omega_{\mathbf q} b^{\dagger}_{\mathbf q} b^{}_{\mathbf q} +  \sum\limits_{i,j=l,u; \mathbf q} \frac{1}{2} \hbar g_{\mathbf q} (b^{\dagger}_{\mathbf q} +  b^{}_{\mathbf q} ) |i\rangle \langle j|. 
\eeq
In the dressed state picture now both diagonal terms, which are of pure dephasing type, and off-diagonal terms, which lead to transitions between the states, appear. In other words, in the dressed state basis, phonons can induce transitions between the states accompanied by phonon emission or absorption, which in the bare states is not allowed. Here, we will consider only low temperatures, and for the sake of simplicity we will set the temperature to $0$~K. At low temperatures phonon absorption processes do not take place, because there are no phonons to be absorbed. Only phonon emission processes occur. This, for example, has been observed experimentally in the state preparation using chirped pulses \cite{luker2012inf,mathew2014sub} or for phonon assisted state preparation \cite{quilter2015pho}.

In the bare state basis, we use a well established fourth order correlation expansion to calculate the dynamics \cite{krugel2005the}, which has been shown to agree well with experiments aiming at the optical state preparation of a QD \cite{reiter2014the,luker2012inf,mathew2014sub,ramsay2010dam} and furthermore is in good agreement with a numerically exact path integral solution \cite{glassl2011lon}. However, due to the pure dephasing type nature of the phonon interaction, we cannot develop a simple analytical model in the bare state basis. Hence, for the analytical solution we will employ the dressed state basis and then transform our findings back to the bare state basis. To validate our analytical equations, we compare both approaches. 

\subsection{QD dynamics}
\label{theo:time}
\begin{figure}[t]
\includegraphics[width=0.8\columnwidth]{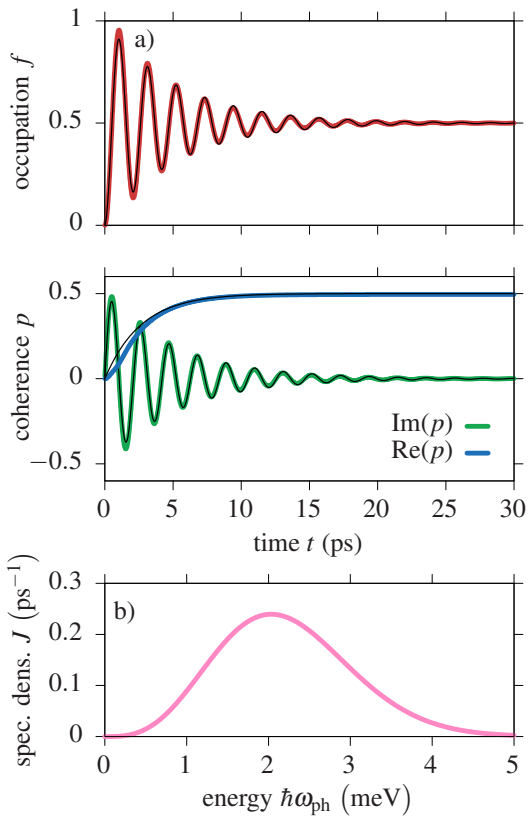}
\caption{(Color online) a) Time evolution of the occupation of the exciton state $f(t)$ and the polarization $p(t)$ calculated numerically within a correlation expansion approach (thick colored lines) and analytical solution according to Eq.~\eqref{eq:eomphonon} (thin black lines). b) Phonon spectral density.}\label{fig:time}
\end{figure}

We start with describing the dynamics of the QD under the influence of phonons and the pump field. On the one hand this is important to understand the time-resolved spectra discussed in the Sec.~\ref{sec:results}, on the other hand we can already test our analytical model.

As a reference, we recall the dynamics of the phonon free case. In the bare state basis, we obtain Rabi oscillations for the dynamics for the occupation $f(t)$ of the exciton state $|2\rangle$ and the polarization $p(t)$ between the two basis states given by
\begin{subequations} \bea
	f(t) &= \varrho_{22} =& \sin^2 \left(\frac{\Omega_R}{2} t \right), \\
	p(t) &= \varrho_{12} =& \frac{i}{2} \sin(\Omega_R t).
\eea \end{subequations}
Note that the occupation of the ground state $g(t)=\varrho_{11}$ always follows from the trace of the density matrix $\varrho$ with $g(t)+f(t)=1$. 

In the dressed state basis the phonon-free dynamics is trivial, because the system is diagonal, i.e. the occupations $\rho^{DS}_{11}$ and $\rho^{DS}_{22}$ are stationary, while the polarization $\rho^{DS}_{12}$ oscillates with the energy splitting of the states $\hbar\Omega_R$. To model the interaction with phonons in the analytical model, we use a semi-phenomenological approach by adding a decay rate $\gamma$ to the dynamics yielding transitions between the dressed states. The analytical model results from a further simplification of the polaron master equation, while the latter was successfully used to describe Rabi rotations \cite{mccutcheon2011gen}. In agreement with the Lindblad formalism, the polarization dephases with half of the decay rate. This leads to the following equations of motion:
\begin{subequations} \bea
	\frac{\partial \varrho^{DS}_{ll}}{\partial t} &=& +  \gamma \varrho^{DS}_{uu}\\ 
	\frac{\partial \varrho^{DS}_{uu}}{\partial t} &=& - \gamma \varrho^{DS}_{uu} \\
	\frac{\partial \varrho^{DS}_{lu}}{\partial t} &=& -i  \,\Omega_R \varrho^{DS}_{lu}  - \frac{\gamma}{2} \varrho^{DS}_{lu} 
\eea \end{subequations}
Setting the initial condition at $t=0$ to $\varrho^{DS}(0)$, the solution of these equations is straight forward:
\begin{subequations} \label{eq:eomphononDS}
\bea
	\varrho^{DS}_{ll}(t) &=& \varrho^{DS}_{ll}(0) + \rho^{DS}_{uu}(0)(1 - e^{-\gamma t}) \\
	\varrho^{DS}_{uu}(t) &=& \varrho^{DS}_{uu}(0) \,\, e^{-\gamma t} \\ 
	\varrho^{DS}_{lu}(t) &=& \varrho^{DS}_{lu}(0) e^{-i \,\Omega_R t}  e^{ -\frac{\gamma}{2} t} 
\eea \end{subequations}
From the dressed states we can go back to the bare states, where the time evolution for the occupation and polarization for the initial condition $\varrho(0)=\varrho_{11}$ reads
\begin{subequations}	\label{eq:eomphonon}
\bea  
		f(t) &=& \frac{1}{2}\left(1- \cos(\Omega_R t) e^{-\frac{\gamma}{2}t} \right) \\
		p(t) &=& \frac{1}{2}\left(1-e^{-\gamma t}\right)	 + i \frac{1}{2} \sin(\Omega_R t) e^{-\frac{\gamma}{2}t}.
\eea \end{subequations} 

In these equation we have two free parameters, namely the Rabi frequency $\Omega_R$ and the decay rate $\gamma$. These are obtained by fitting the analytical solution to the numerical solutions,  while the latter has been calculated using realistic QD parameters with confinement length for the electrons of $3$~nm and material parameters typical for GaAs \cite{krummheuer2002the,wigger2014ene}. To achieve maximal coupling to the phonons, we set the Rabi frequency to $\Omega_R=3$~ps corresponding to a period of $T_R=2\pi/\Omega_R=2.1$~ps, such that the energy of $\hbar\Omega_R\approx2$~meV agrees with the maximum of the phonon spectral density \cite{wigger2014ene}, as shown in Fig.~\ref{fig:time}~b).

The dynamics of the exciton occupation $f$ and the polarization $p$ are shown in Fig.~\ref{fig:time}~a). The occupation performs Rabi oscillations, which are damped rather quickly by the interaction with the phonons due to the efficient coupling. For higher/lower Rabi frequencies, the damping would be less pronounced and the Rabi oscillations would last longer. The imaginary part of the polarization also shows a damped oscillatory behavior with the final stationary value of $\mbox{Im} (p(\infty))=0$, while the real part increases to about $\mbox{Re}(p(\infty)) \approx 0.5$. The stationary values can be approximately described by the dressed state basis, where the system has relaxed to the lower dressed state with $|l\rangle = \frac{1}{\sqrt{2}}(|1\rangle+2\rangle)$ \cite{glassl2011lon}. 

The thin black lines shows the analytical solution from Eq.~\eqref{eq:eomphonon}. There is a very good agreement between numerical and analytical solution confirming the validity of our model. Only for very short times, a noticeable difference between the analytical solution and the numerical solution is seen in the real part of the polarization starting with non-zero and zero slope, respectively. This is due to the difference between the Markovian and non-Markovian treatment. By fitting, the parameters for the analytical solution are obtained to $\Omega_R=2.996$~meV and $\gamma=0.377$~ps$^{-1}$. The value for $\gamma$ agrees well with the maximal value of spectral density at $\hbar\omega_{max}=2$~meV with $J(\hbar\omega_{max})=0.24$~ps$^{-1}$ corresponding to a rate calculated via Fermi's Golden rule $\gamma=\frac{\pi}{2}J(\omega)=0.377$~ps$^{-1}$ \cite{barth2016fas}. In the following, for the calcuations with phonons we will make use of these parameters.

\subsection{Calculation of the probe spectrum}
\label{theo:spec}
Now we want to analyze the time-resolved signals in a pump-probe set-up, where in our case the pump pulse is the continuous light field switched on instantaneously, while the probe pulse is an ultra-short, weak pulse acting after a time delay $\tau$ after the switch-on of the pump pulse. To experimentally extract the probe signal in spatially extended systems like quantum wells a spatial separation via the propagation direction is performed \cite{mukamel2000mul}, however, for a single QD this is not possible. Instead, for single dot experiments the system is excited by a series of laser pulses with slightly different frequencies. By heterodyne detection the frequency component corresponding to different optical signals can be extracted, as has successfully been implemented to study of FWM of single QDs \cite{kasprzak2013coh,mermillod2016dyn}. To obtain a differential pump-probe spectrum typically two runs are performed, one with and one without pump pulse, and the difference of the signals is considered \cite{sotier2009fem}. In this case, however, the coherence between pump and probe arising due to the overlap of the pulses is lost.

In the simulation, the probe signal is modelled by adding a phase $\varphi$ to the probe pulse and then sorting out the desired signals from the polarization. In the numerical solution, the probe pulse is modelled as a Gaussian pulse with a width of $10$~fs. The probe polarization $p^P$ is obtained by filtering the phase of the probe pulse using a Fourier expansion \cite{haug1996qua,axt2004fem,reiter2013opt}. In the analytical solution the probe pulse is assumed to be a $\delta$-pulse with a phase $\varphi$. Its action can be described by multiplication of matrices simulating the action of the pulse \cite{vagov2002ele,axt2005pho}. After the probe pulse the polarization contains contributions with different phases, resembling the different frequency components. From the polarization only the components proportional to  $e^{-i \varphi}$ are selected for the probe signal yielding the probe polarization $p^{P}$, while other components resemble, for example, the FWM signal \cite{mermillod2016dyn}.

From the probe polarization $p^{P}$ the absorption spectrum is obtained by a Fourier transform and taking the imaginary part via
\beq \label{eq:FTgamma}
	\alpha(\omega) =\mbox{Im}(  {\cal{FT}}(p^{P}) ) \propto \mbox{Im} \left(\int\limits_{-\infty}^{\infty} p^{P}(t) \,  e^{ i \omega t } dt \right).
\eeq
In principle, to obtain the spectrum one has to divide the Fourier transform of the polarization by the Fourier transform of the electric field, which we omit here, because in the limit of ultra-short or $\delta$-pulses the Fourier-transform of the electric field can be approximated by a constant.

\section{Probe spectra}
\label{sec:results}
\subsection{Without phonons}\label{results:ohnePh}

\begin{figure}[t]
\includegraphics[width=\columnwidth]{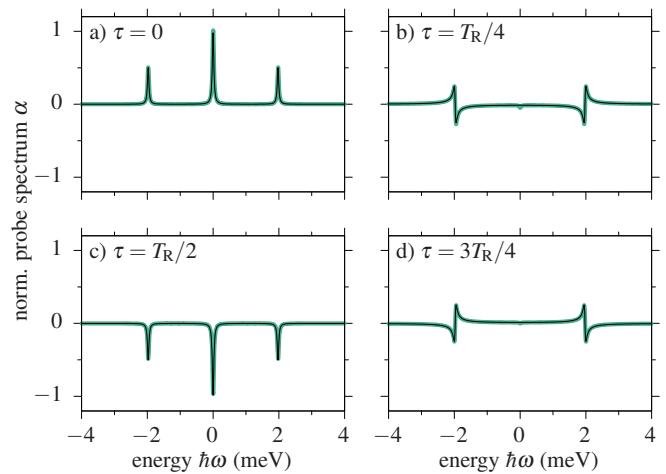}
\caption{(Color online) Normalized probe spectra for different time delays a) $0$, b) $T_R/4$, c) $T_R/2$ and d) $3T_R/4$. The green line depicts the numerical solution and the thin black line the analytical solution. \label{fig:ohne}}
\end{figure}

As a reference, we first study the probe spectra without phonons. The continuous light field sets in at time $t=-\tau$, where $\tau$ denotes the delay between the switch-on of the continuous field and the probe pulse. When the probe pulse acts, the system is in an arbitrary state expressed by the density matrix $\varrho(0)=\{ \varrho_{ij}^{(0)} \}$. Without phonons, the polarization does not dephase at all. To add a dephasing, which simulates a finite lifetime, we multiply the polarization with an exponential $e^{-\Gamma t}$ having the dephasing rate $\Gamma$. Then the probe polarization calculates to 
\bea \label{eq:pp_ohne}
 p^{P}(t) &=&  \Theta(t) \frac{i}{4} \big[ 2 ( \varrho_{11}^{(0)} - \varrho_{22}^{(0)})  \\ \nonumber
 		&+&   e^{ i \Omega_R t} (\varrho_{11}^{(0)} + 2 \varrho_{21}^{(0)} - \varrho_{22}^{(0)}) \\ \nonumber
 		&+&  e^{- i \Omega_R t} (\varrho_{11}^{(0)} - 2 \varrho_{21}^{(0)} - \varrho_{22}^{(0)}) \big] e^{-\Gamma t}
\eea
and the spectrum is:
\bea \label{eq:abs_ohne}
	&& \alpha(\omega) = \frac{1}{4\sqrt{2\pi}} \left\{
	 		     \frac{2\Gamma \left[\varrho_{11}^{(0)}-\varrho_{22}^{(0)} \right]}{\Gamma^2 + \omega^2} \right.  \\ \nonumber
	&& 		     + \frac{\Gamma\left[\varrho_{11}^{(0)} + 2 \mbox{Re}(\varrho_{21}^{(0)}) - \varrho_{22}^{(0)}
	 		    		+(-\omega + \Omega_R) 2\mbox{Im}(\varrho_{21}^{(0)})\right]}{\Gamma^2 + (\omega-\Omega_R)^2} \\ \nonumber
	&& 		    \left. + \frac{\Gamma \left[\varrho_{11}^{(0)} - 2 \mbox{Re}(\varrho_{21}^{(0)}) - \varrho_{22}^{(0)}
	 		   		+ (\omega + \Omega_R) 2\mbox{Im}(\varrho_{21}^{(0)})\right]}{\Gamma^2 + (\omega+\Omega_R)^2} \right\} 
\eea	
Already in Eq.~\eqref{eq:abs_ohne} we see that up to three peaks appear in the spectrum, where the denominators are minimal, at $\hbar\omega=0$ and at the Rabi energies $\hbar\omega=\pm\hbar\Omega_R$. Note that due to the rotating frame, $\hbar \omega=0$ corresponds to the transition energy $E_{12}$. The strength of the peaks is determined by the occupations and the polarization at time $t=0$, while the width for all peaks is given by the decay rate $\Gamma$. The spectra for the time delays $\tau=0, T_R/4, T_R/2$ and $3T_R/4$ are shown in Fig.~\ref{fig:ohne}. All spectra are normalized to the amplitude of the middle peak at $\tau=0$. Note that the spectra for $\tau \to \tau+n T_R$ with $n\in \{0,1,2,3..\}$ are the same. The first time delay at $\tau=0$ corresponds to the time when the system is in its ground state and only the density matrix element $\varrho_{11}^{(0)}=1$ is non-zero. Three peaks come up, while the middle peak is twice as high as the side peaks. This is analogous to the Mollow triplet \cite{gerry2005int} and can be understood by looking at the transitions between the dressed states in Fig.~\ref{fig:scheme}: The transition with energy $E_{12}$ appears twice, while the transitions with the energies $E_{12}\pm\hbar\Omega_R$ appear only once. The positive amplitudes can be related to the fact that the system is in the ground state and is thus absorbing. After half a period at $\tau=T_R/2$~ps, the system is in the excited state, such that the density matrix has only one entry with $\varrho_{22}^{(0)}=1$. Again, the spectrum consists of three peaks, but this time with negative amplitudes, which is expected because in the excited state the system should exhibit gain. At the intermediate times at $\tau=T_R/4$ and $\tau=3T_R/4$ the ground and the excited state are half occupied with $\varrho_{11}^{(0)}=\varrho_{22}^{(0)}=1/2$, while the coherence is $\varrho_{12}^{(0)}=\pm i/2$. Therefore, in the spectrum the middle peak, which only depends on the difference in the occupations, vanishes. The side peaks show a dispersive behavior due to the purely imaginary polarization. For intermediate times the spectrum evolves smoothly from one form into the other. The results were obtained with the analytical solution (thin black lines) and the numerical solution (thick green lines) yielding the same spectra.

\begin{figure}[t]
\includegraphics[width=\columnwidth]{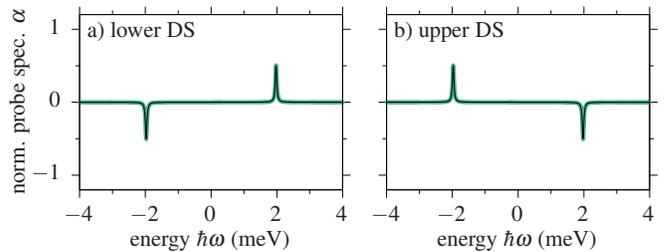}
\caption{(Color online) Normalized probe spectra for system being in a) the lower and b) the upper dressed state. The green line depicts the numerical solution and the thin black line the analytical solution.\label{fig:ohneDS}}
\end{figure}

It is also instructive to study the probe spectra when the system is in one of the dressed states $|l\rangle$ or $|u\rangle$ shown in Fig.~\ref{fig:ohneDS}. For both dressed states the occupations are equal with  $\varrho_{11}^{(0)}=\varrho_{22}^{(0)}=1/2$. In analogy to the superposition state at $\tau=(2n+1)T_R/4$ during the dynamics, the middle peak in the spectrum vanishes. In contrast, the polarization of the dressed states is real with $\varrho_{12}^{(0)}=\pm 1/2$, where the $+$ belongs to the lower dressed state and the $-$ to the upper dressed state. Correspondingly, we find the side peaks exhibit one absorptive and one gain signature. The signs of the amplitudes of the peaks can be understood from looking at the dressed state picture in Fig.~\ref{fig:scheme}. Let us concentrate on the spectrum resulting from the lower dressed state: From the lower dressed state the system emits light with the energy $E_{12}-\hbar\Omega_R$ ending up in the upper dressed state of the lower replica. Accordingly, the peak at $-\hbar\Omega_R$ has a negative sign in the lower dressed state spectra. Likewise the system absorbs light at the energy $E_{12}+\hbar\Omega_R$ yielding a positive amplitude at $\hbar \Omega_R$. Though also transitions between the replica from lower to lower dressed state are allowed, here absorption and gain cancel each other. The spectrum for the upper dressed state is reversed and can be explained likewise. 

\subsection{With phonons} \label{results:mitPh}
\begin{figure}[ht]
\includegraphics[width=0.7\columnwidth]{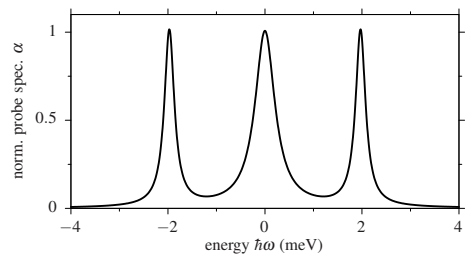}
\caption{Normalized probe spectra for $\tau=0$ from the analytical solution. \label{fig:mittau0}}
\end{figure}

Now we include phonons, which lead to a fast damping of the Rabi oscillations as discussed above (see Fig.~\ref{fig:time}~a)). We consider the case where phonon induced dephasing is much faster than a dephasing induced by other sources, hence, we set the rate $\Gamma$ to zero. To calculate the action of the probe pulse we use the matrix multiplication method from Ref.~\cite{vagov2002ele,axt2005pho} and calculate the subsequent dynamics again in the dressed state picture. The probe polarization calculates to
\bea \label{eq:p_mit}
	 p^{P}(t) &=& \Theta(t) \frac{i}{4} \left[ ( \varrho_{11}^{(0)} - \varrho_{22}^{(0)}) e^{-\gamma t} \right. \\ \nonumber
	 &+&   e^{ i \Omega_R t} e^{-\frac{\gamma}{2} t} (\varrho_{11}^{(0)} + 2 \varrho_{21}^{(0)} - \varrho_{22}^{(0)}) \\ \nonumber
	 &+&  \left.  e^{ -i \Omega_R t} e^{-\frac{\gamma}{2} t} (\varrho_{11}^{(0)} - 2 \varrho_{21}^{(0)} - \varrho_{22}^{(0)}) \right].
\eea
Already from this equation we can see a difference of the phonon influence in comparison to multiplication with a single rate as done in Eq.~\eqref{eq:pp_ohne}. The first term $\sim( \varrho_{11}^{(0)} - \varrho_{22}^{(0)})$ now decays with the rate $\gamma$, while the other two terms, which have an oscillatory time dependence with $e^{\pm i \Omega_R t} $, decay with only half the phonon rate $\gamma/2$. Inserting the time behavior of the occupations from Eq.~\eqref{eq:eomphononDS}, the spectrum for $\tau>0$ evaluates to

\bea 
&&	\alpha(\omega) = \frac{1}{8\sqrt{2\pi}} \left\{
	 		    \frac{4\gamma \, \cos( \Omega_R \tau)\, e^{-\frac{\gamma}{2}\tau} }
	 		      {\gamma^2 + \omega^2}  \right. \\ \nonumber
&&+	 		  \frac{\gamma - \gamma e^{-\gamma\tau}+e^{-\frac{\gamma}{2}\tau}\left[\gamma\cos(\Omega_R \tau) + 2(-\omega+ \Omega_R) \sin(\Omega_R \tau) \right] }
	 		  {\left(\frac{\gamma}{2}\right)^2 + (\omega-\Omega_R)^2} \\ \nonumber
&&+	 		   \frac{-\gamma + \gamma e^{-\gamma\tau}+e^{-\frac{\gamma}{2}\tau}\left[\gamma\cos(\Omega_R \tau) + 2(\omega+ \Omega_R) \sin(\Omega_R \tau) \right] }
	 		   	{\left(\frac{\gamma}{2}\right)^2 + (\omega+\Omega_R)^2}
	 		    \bigg\}.
\eea 

A special case is the spectrum at $\tau=0$ shown in Fig.~\ref{fig:mittau0}, when the system is in the ground state with $\varrho_{11}=1$ and no phonon damping has taken place yet. Nevertheless, the phonons influence the probe spectrum and we obtain a different spectrum as without phonons from Fig.~\ref{fig:ohne}. All three peaks now have the same height, but different widths according to the different exponentials found in the polarization in Eq.~\eqref{eq:p_mit}. The ratio of 2:1 of the amplitudes between the middle and the side peaks is found in the area under the peaks. This result is in agreement with calculations using a quantized light field \cite{vagov2014com}.

\begin{figure}[t]
\includegraphics[width=\columnwidth]{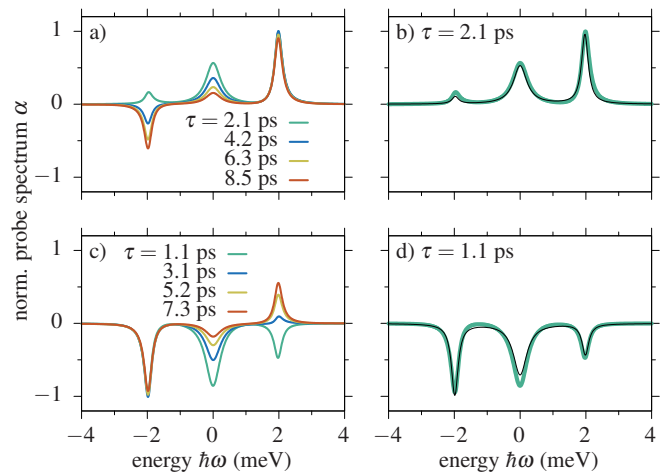}
\caption{(Color online) Normalized probe spectra including phonons for (a) $\tau$ at a minimum of the exciton occupation and (c) $\tau$ at the maximum of the exciton occupation. (b) and (d) show two examples to compare the numerical (thick green lines) with the analytical solution (thin black lines). \label{fig:mitspectren}}
\end{figure}

Next, we study the spectra for $\tau>0$ showing the spectra at different $\tau$ in Fig.~\ref{fig:mitspectren} obtained from the numerical solution. All spectra are normalized to the maximum value appearing at $\tau=0$. In Fig.~\ref{fig:mitspectren}~a) the delays $\tau$ are chosen to coincide with the minima of the exciton occupation in Fig.~\ref{fig:time}, i.e., a maximum of the ground state occupation. At $\tau=2.1$~ps we find three peaks with positive but different amplitudes. When the time delay increases, the middle peak at $\hbar\omega=0$ vanishes. The right peak at $\hbar\omega=\hbar\Omega_R$ remains almost unchanged as a function of the time delay. In contrast, the left peak at $\hbar\omega=-\hbar\Omega_R$ first diminishes and then its amplitude changes its sign from positive to negative, regaining strength. In Fig.~\ref{fig:mitspectren}~c) the time delays are shifted by half a period to be at maxima of the exciton occupation. Again, the middle peak at $\hbar\omega=0$ decreases for increasing time delay. In contrast, for maxima of the exciton occupation the left peak at $\hbar\omega=-\hbar\Omega_R$ remains unchanged, while the right peak at $\hbar\omega=\hbar\Omega_R$ changes its sign. Both spectra approach a two-peak spectrum with a gain peak at $\hbar\omega=-\hbar\Omega_R$ and an absorptive peak at $\hbar\omega=\hbar\Omega_R$. This behavior reflects that due to the phonon interaction the system relaxes into the lower dressed state, where a two peak spectrum as in Fig.~\ref{fig:ohneDS}(a) is found. 

The same behavior is found by the analytical solution. As an example we compare the spectra for $\tau=2.1$~ps and $\tau=1.1$~ps in Fig.~\ref{fig:mitspectren}(b) and (d), respectively. The analytical solution (thin black line) agrees again well with numerical solution (thick green line). For the numerical solution it is difficult to calculate the probe polarization at exactly $\tau=0$ correctly due to the finite pulse length, thus, in Fig.~\ref{fig:mittau0} only the analytical solution is shown.

\begin{figure}[t]
\includegraphics[width=\columnwidth]{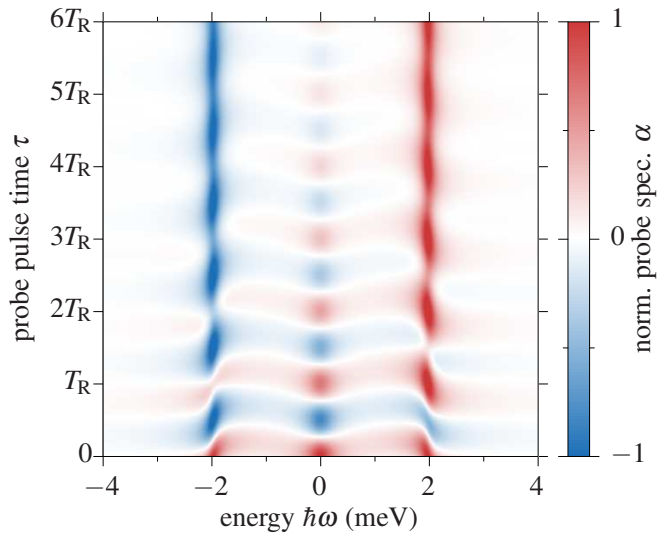}
\caption{(Color online) Color plot of the normalized probe spectra as function of delay time $\tau$ calculated by the analytical model. \label{fig:mit3D}}
\end{figure}

The dynamics is summarized in the color plot of the spectrum as s function of time delay up to $6T_R$ in Fig.~\ref{fig:mit3D}, where we have used the analytical equation to obtain the results. In the first few periods we find the oscillatory behavior between positive and negative spectra as discussed in Sec.~\ref{results:ohnePh}. Due to the strong phonon coupling the dephasing is quite fast and already after a few cycles the Rabi oscillations get damped (cf. Fig.~\ref{fig:time}). If the phonon coupling would be weaker, the oscillations could also be seen on longer time scales. For larger time delays in Fig.~\ref{fig:mit3D} we see that the amplitude of the middle peak vanishes, while the left peak becomes only negative and the right peak becomes positive. We point out that the relaxation process cannot be described by a dephasing of the polarization only, which would result in the same final occupations, but a different polarization, and hence a different spectrum.

\subsection{Interplay of phonons and decay} \label{sec:gammagamma}

\begin{figure}[t]
\includegraphics[width=\columnwidth]{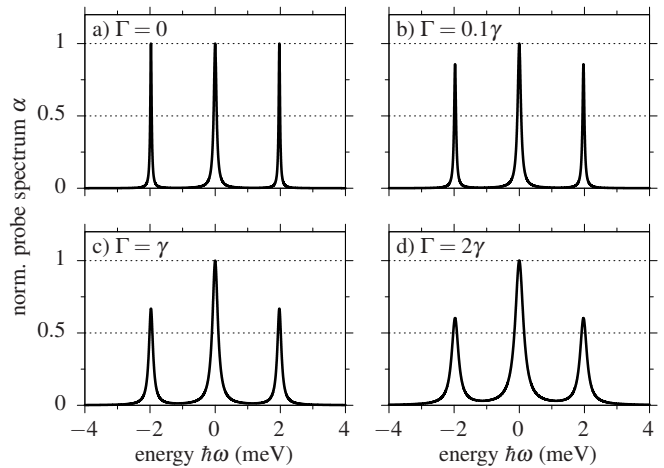}
\caption{Probe spectra at $\tau=0$ calculated from the analytical model including phonon coupling via $\gamma$ and the dephasing rate $\Gamma$ with a) $\Gamma=0$, b)$\Gamma=0.1\gamma$, c) $\Gamma=\gamma$ and d) $\Gamma=2\gamma$. \label{fig:gammagamma}}
\end{figure}

Finally, we want to analyze the interplay between the phonon influence and the effect of a dephasing rate $\Gamma$ as introduced for the phonon free case in Sec.~\ref{results:ohnePh}. Such an interplay is important for smaller electron-phonon coupling strengths, where the polarization is less damped by phonons and other dephasing mechanisms will set in. For this purpose, we use the analytical model to calculate the probe polarization and multiply it with an exponential $e^{-\Gamma t}$ with the rate $\Gamma$. To see the effect, we use a much lower value for the phonon rate of $\gamma=0.05$~ps$^{-1}$ to have a sharp spectrum and increase $\Gamma$ from $0$ to $0.1$~ps$^{-1}$ (from $0$ to $2\gamma$) for the spectra shown in Fig.~\ref{fig:gammagamma}. All spectra are normalized to the amplitude of the middle peak at $\omega=0$. The arrival time of the probe pulse is set to $\tau=0$. For $\Gamma=0$ in Fig.~\ref{fig:gammagamma}~a), we find the same behavior as in Fig.~\ref{fig:mittau0} with all peaks having the same height. When we increase $\Gamma$ in Fig.~\ref{fig:gammagamma}~b) and c), we see that the side peak amplitude decreases in comparison to the amplitude of the middle peak. For $\Gamma=2\gamma$ in Fig.~\ref{fig:gammagamma}~d) it almost approaches the 2:1 ratio we have found in the phonon-free case. Likewise the width of the side peaks increases to be similar to the one of the center peak. In total, the spectrum approaches the phonon-free case. Note that in all cases the area below the middle peak is twice the area below one side peak.

We close the discussion by a comparison of the time-resolved signals presented here with the Mollow triplet observed in resonance fluorescence. Resonance fluorescence is a steady state signal obtained by looking at the photons emitted by spontaneous decay resulting in a spectrum with three peaks. Also for a QD embedded in a cavity the Mollow triplet has been observed \cite{ulhaq2010lin,ulhaq2013det,wei2014tem,unsleber2015obs,fischer2016sel} in contrast to the two-peak spectrum predicted here for the steady state. The main difference between these observations and the present study is caused by the spontaneous decay. If we consider the case with only spontaneous emission and no phonons, the system relaxes to the ground state, which is a superposition of upper and lower dressed state resulting in the Mollow triplet. When both spontaneous emission and phonons are included, the system ends up in a statistical mixture of upper and lower dressed state \cite{gauger2010hea} and hence a three peak spectrum should be observed in the final state. Here we are in the opposite limiting case, where the phonon relaxation is much faster than the spontaneous decay. Note that in fact spontaneous emission has been neglected here. Accordingly the system can relax to the lower dressed state and will not be promoted back to the upper dressed state by spontaneous emission. Such a relaxation in the lower dressed state takes place, e.g., for the phonon damping of Rabi rotations of a single QD \cite{ramsay2010dam}.

\section{Conclusions}\label{sec:concl}  
In conclusion, we have studied the time-resolved probe spectra of a continuously driven QD under the influence of phonons. Using a semiclassical rate equation model, we derived analytical equations for the spectra, which we compared to numerical studies within a well established correlation expansion approach. We showed that the phonons modify the three-peak spectrum, such that the 2:1 ratio in the amplitudes of the peaks is found in the areas under the peaks. Studying the time evolution of the spectra, two regimes were identified: On the time scale of the Rabi period, the spectra oscillate between peaks with positive and negative amplitude, while on the time scale of the phonon damping, the central peaks vanishes and the phonon-induced relaxation to the lower dressed state is monitored in the probe spectra, which for long times consists only of the two side peaks. The results are easily transferable to an excitation with a finite pulse, e.g. a Gaussian pulse of a few ps, as pump pulse. In our study we have derived simple analytic equations and discussed the impact of phonons on optical signals of a QD.

\section{ACKNOWLEDGMENTS}

I am grateful for many useful discussions with my colleagues, in particular with Tilmann Kuhn, Daniel Wigger and Thomas Papenkort.


%

\end{document}